\documentstyle[12pt]{article}
\advance\textheight by \topskip
\begin{document}
\rightline{SU-ITP-92-27}
\rightline{\today}
\vspace{1.2cm}
\begin{center}
{\large\bf  DILATON BLACK HOLES NEAR THE HORIZON}

\vskip .9 cm

{\bf Renata Kallosh \footnote {On leave  from: Lebedev
Physical Institute, Moscow. \  E-mail address:
kallosh@physics.stanford.edu}
 and Amanda Peet  \footnote{E-mail address: peet@
slacvm.slac.stanford.edu} }
 \vskip 0.1cm
Physics Department, Stanford University,
Stanford    CA 94305
\end{center}
\vskip .6 cm
\centerline{\bf ABSTRACT}
\vspace{-0.7cm}
\begin{quote}

Generic $U(1)^2$ 4-d black holes
with  unbroken $N=1$ supersymmetry are shown to tend to
a Robinson-Bertotti type geometry  with a linear dilaton and
doubling of unbroken supersymmetries near the
horizon.  Purely magnetic dilatonic black
holes, which have unbroken $N=2$ supersymmetry,
behave near the horizon as a 2-d linear dilaton
vacuum $\otimes \,  S^2$. This geometry is invariant
under 8 supersymmetries,
i.e. half of the original $N=4$ supersymmetries are  unbroken.
 The supersymmetric positivity bound,
which requires the mass of the 4-d dilaton
black holes to be greater than or equal to the central
charge, corresponds to
positivity of mass for a class of stringy 2-d  black holes.

\end{quote}

\normalsize
\newpage

1. The evaporation of stringy
$U(1)^2$ charged black holes \cite{GM}, \cite{GHS} may be
understood as the
process of restoration of supersymmetry \cite{US}.  It is likely that
the
endpoints of the process of evaporation for  charged dilatonic black
holes are
stable remnants which are zero temperature extreme black holes with
some
unbroken supersymmetry.
Those black holes have the minimum possible
mass, $M = \frac{1}{\sqrt
2} \,( |Q|+|P|)$. It was shown in \cite{US} that
the extreme solutions  saturate the
 supersymmetry bound of $N=4,  d=4$ supergravity, or dimensionally
 reduced superstring theory. When both electric and magnetic charges
are
 present, only one bound of $N=4$ supersymmetry is saturated and the
corresponding solution has unbroken $N=1$ supersymmetry. In the case
of only
 electric or only magnetic charges, i.e. just $U(1)$ black holes,
both
supersymmetry bounds of  $N=4$ supersymmetry are saturated and these
black holes have unbroken $N=2$ supersymmetry \cite{US}.

Several properties of extreme dilatonic black holes were investigated
in \cite{US}. The purpose of
this paper is to investigate the properties of extreme
supersymmetric stringy black holes in the neighborhood of the
 horizon, to find what kind of a geometry they tend to and what
happens with unbroken supersymmetries in those geometries.
We would also like to understand the relation to 2-d dilatonic black
holes and
whether the investigations of the supersymmetry and of the geometry
near the
horizon of the 4-d black hole may be an important factor for
understanding the late stages of evaporation of 2-d black holes.

 We will start by describing the behavior of the familiar extreme
Reissner-Nordstr\"om black hole near the horizon \cite{GWG}, where
the
geometry becomes that of the Robinson-Bertotti solution and has twice
as many
supersymmetries as the extreme black hole.

Then we will study properties of electric-magnetic dilaton black
holes near the
horizon. The difference with the Reissner-Nordstr\"om black hole
near the horizon will show up in the existence of a non-constant
dilaton, the
linear dependence on the coordinate    being proportional to the
dilaton
charge. We will find, however, that the essential
properties of the geometry at the horizon are very close to those of
the Robinson-Bertotti
solution and also that doubling of supersymmetries takes place.

The geometry near the horizon and the supersymmetry properties of
purely
magnetic extreme black holes is the next topic to be studied. The
fact
that near  the horizon these solutions form a geometry which is the
direct
product
of a 2-d linear dilaton vacuum and a 2-sphere of constant curvature
has been
established before \cite{GS}. It was argued that, at scales where the
2-sphere
radius, which is  proportional to the magnetic charge $P$, can be
neglected,
the
black hole physics may be described by a 2-d effective theory. This
is  one of
the
 reasons for the recent interest in two-dimensional black holes.  It
has also
been established
 before \cite{US} that the extreme magnetic dilaton black hole has
 8 unbroken parameters of $N=2$ supersymmetry, i.e. one half of the
original
16 parameters of $N=4$. In this paper
we analyze what happens with unbroken supersymmetries of the extreme
black
hole
at the horizon. We find that the solution (a direct product of the
2-d linear
dilaton vacuum and a 2-sphere of constant curvature) has 8 unbroken
parameters of $N=2$
supersymmetry, exactly as does the total geometry of the extreme
black
hole; there is no doubling of the supersymmetries to the maximum
possible $N=4$.  Finally, the relation between the
supersymmetric positivity bound $M \geq \frac{1}{\sqrt
2} \,|P|$ in 4-d and positivity of mass in 2-d is exhibited.

\vskip 0.6 cm
2. The extreme Reissner-Nordstr\"om black hole near the horizon has
been
investigated before \cite{GWG}.
There are several properties of the extreme Reissner-N\"ordstrom
black hole which make it interesting.  One is that it has zero
temperature, so
is
 stable to emission of Hawking radiation.  The extreme
Reissner-Nordstr\"om
configuration also possesses unbroken $N=1$ supersymmetry
\cite{Gibbons},
\cite{GWG} when viewed as a bosonic configuration of $N=2,~d=4$
supergravity.
 A further property is that near its horizon the extreme
Reissner-Nordstr\"om
geometry  asymptotes to a Robinson-Bertotti geometry, which is a
maximally
supersymmetric \cite{GWG} (and homogeneous)
configuration of $N=2,~ d=4$ supergravity.
To see this, consider the
extreme  Reissner-Nordstr\"om (RN) metric in isotropic coordinates
$\left\{ x^i
\right\}$
\begin{equation} ds^{2}_{RN} =
V_{RN}^{2}dt^{2}-V_{RN}^{-2}d{\vec{x}}^{2}\ ,
\end{equation}
where $|\vec{x}| = \rho$ and
\begin{equation}
V_{RN}^{-1}(\rho) = 1 + \frac{M}{\rho}\ .
\end{equation}
The Maxwell field is
\begin{equation}
F_{RN} = \pm d V_{RN} \wedge dt\ ,
\end{equation}
and $M=|Q|$.  Near the horizon $\rho \rightarrow  0$, the function
$V_{RN}^{-1}
\rightarrow  \frac{M}{\rho} = V_{RB}^{-1}$, i.e.  the extreme
Reissner-Nordstr\"om metric tends to the Robinson-Bertotti  (RB)
metric there,
and the Maxwell field tends to that of the RB configuration:
\begin{equation} ds^2_{RB} =
\frac{{\rho}^2}{M^2} \,  dt^2 -\frac{M^2}{\rho^2}
 \, d \rho^2 - M^2  \, d \Omega^2 \ ,
\qquad F_{RB}  = \pm{1 \over M}d \rho \wedge dt\ .
\label{BR}\end{equation}

Since the extreme Reissner-Nordstr\"om
geometry admits $N=1$ supersymmetry and  the \\Robinson-Bertotti
geometry
admits $N=2$ supersymmetry, one may call this phenomenon ``doubling
of
supersymmetries near the black hole horizon". To explain this
doubling of
supersymmetries, let us consider the supersymmetric transformation of
the {\it
gravitino field strength }\cite{K},
\begin {equation}\label{gravitino} \delta\Psi_
{ABC I} = C_{ABCD}^{+} \epsilon^{D+}_I  +
C_{ABCD}^{-}  \epsilon^{D-}_I  \ .
\end{equation}
In eq.(\ref{gravitino}), 2-dimensional spinor notation is used.
The supersymmetry parameters $\epsilon^{\pm}_I, I = 1,2,$ are defined
in eqs.
(37) of ref. \cite{K}, and $C^{\pm}$ are the following combinations
of the Weyl
($C_{ABCD}$) and Maxwell ($F_{CD} $) spinors:
 \begin{equation}\label{cpm}
C_{ABCD}^{\pm}  \equiv C_{ABCD} \pm  \nabla_{AB'} \, F_{CD} \,
V^{-1} \,
K^{B'}_{\ B}  \ ,
\end{equation}
where $K^{B'}_{\ B} $ is the Killing vector. For the extreme
Reissner-Nordstr\"om black hole, we have either $C_{ABCD}^+=
0$ or $C_{ABCD}^- = 0$  (depending on the sign of the charge $Q$),
which is a
relation between the Weyl spinor and the derivative of the Maxwell
spinor. In
the first case the unbroken $N=1$ supersymmetry parameter of the
extreme Reissner-Nordstr\"om black hole is $\epsilon^{+}_I $ , in the
second
case it is $\epsilon^{-}_I $.

Near the horizon, the Reissner-Nordstr\"om geometry becomes that of
the
Robinson-Bertotti solution, which is {\it conformally flat and has a
covariantly constant Maxwell field},
\begin{equation}
C_{ABCD}=0 ,\qquad \nabla_{AB'} \, F_{CD}=0\ .
\label{conf}
\end{equation}
Thus, both combinations of the Weyl and Maxwell spinors which enter
the
supersymmetry variation of the gravitino field strength
(\ref{gravitino}) are
vanishing in this geometry. This property ensures that all 8
parameters
($\epsilon^{+}_I $ and $\epsilon^{-}_I $) of the original $N=2$
supersymmetry
are unbroken  in the  Robinson-Bertotti background.

We see that the sign of the charge controls which supersymmetries are
unbroken
in the Reissner-Nordstr\"om solution, but that no such phenomenon
occurs for
the Robinson-Bertotti solution where {\it all} supersymmetries are
unbroken due
to the separate vanishing of both terms making up the generalized
Weyl
curvatures $C^{\pm}$ in eq. (\ref{cpm}).

Notice also that the extreme Reissner-Nordstr\"om metric is
 asymptotically Minkowskian, since at infinity only the constant term
in
$V_{RN}^{-1}$ survives.
Minkowski space is also a maximally supersymmetric geometry.
Therefore, in
some sense, the
extreme RN solution may  be thought of as a soliton; it interpolates
between
two candidate vacua for  $N=2, d=4$ supergravity \cite{Gibbons}.

A natural question now arises -- how much of this behavior carries
over to
extreme dilaton black holes ?
\vskip 0.6 cm
3. Doubling of supersymmetries near the horizon for extreme
electric-magnetic dilaton black holes will be explained in what
follows.

The  action we will use is the part of the $SO(4)$ version of the
$N=4, d=4$
supergravity action without axion,
 \begin{equation}\label{so4action}
 I =\frac{1}{16\pi} \int d^4x\,\sqrt{-g} \Biggl[ -R +
2\, \partial^\mu \phi \cdot \partial_\mu \phi- \left(
{\mbox{e}}^{-2\phi}F^{\mu\nu}F_{\mu\nu} + {\mbox{e}}^ {2\phi}\tilde
G^{\mu\nu}\tilde G_{\mu\nu}\right) \Biggr]\ ,
\end{equation}
where  $\tilde G_{\mu\nu}$ is related to the
non-dually rotated field $G_{\mu\nu}$ as follows
 \begin{equation}\label{dual}
\tilde G^{\mu\nu} = {\frac{i}{2}}
\frac {1}{\sqrt{-g}}\, e^{-2\phi}\,
\epsilon^{\mu\nu\lambda\delta}\,
G_{\lambda\delta}\ . \end{equation}
 All notation is that of \cite{US}.
For extreme  supersymmetric dilatonic black holes, the fields are
built out of
two functions $H_1$ and $H_2$ \cite{US} :
 \begin{eqnarray}
ds^{2} &=&  e^{2U}dt^{2}-e^{-2U}d\vec{x}^{2} \ ,\nonumber\\
A = \psi dt  &, &\qquad \tilde B = \chi dt  \ , \nonumber\\
F =  d \psi \wedge dt &, &\qquad  \tilde G = d \chi \wedge dt \ ,
\nonumber\\
e^{-2U} = H_1 H_2  &, &\qquad
e^{2\phi} = H_2/ H_1  \ ,\nonumber\\
\sqrt{2}\, \psi = \pm H_1^{-1} &,&\qquad  \sqrt{2}\,\chi
=\pm H_2^{-1} \ ,
\label{anz}\end{eqnarray}
where the condition on the functions $H_1,H_2$ is that they be
harmonic,
\begin{equation} \label{Hcond}
\partial_i\partial_i H_1 =0\ , \qquad \partial_i\partial_i H_2=0 \,
{}. \end{equation}
We have used isotropic coordinates $\left\{ x^i \right\}$, where
$\rho^2 = x^i x^i$ and
\begin{equation}
H_1 = {\mbox{e}}^{-\phi_0} (1+ \frac{\sqrt {2 }|Q|}{\rho}) \ , \qquad
H_2 = {\mbox{e}}^{+\phi_0} (1+\frac{\sqrt {2}|P|}{\rho}) \ .
\label{bh}\end{equation}
The mass $M$ and dilaton charge $\Sigma$ are related to the $U(1)$
electric $Q$
and magnetic $P$ charges as
\begin{equation}
M =\frac{ |P| + |Q|}{\sqrt {2 }}\ , \qquad  \Sigma = \frac{ |P| -
|Q|} {\sqrt {2}} \, \, .
\end{equation}

It was shown in \cite{US} that the bosonic background
(\ref{anz}), (\ref{Hcond}) admits  supercovariant
Killing spinors of $N=4, d=4$ supergravity, i.e. that there exist
non-trivial solutions of the equations
\begin{equation}
\delta \Psi _{\mu I }(\epsilon )=
\delta \Lambda _I(\epsilon ) = 0 \ , \qquad I = 1,2,3,4,
\label{kil}\end{equation}
 describing the supersymmetry variation of 4 gravitinos and 4
dilatinos
 in the background (\ref{anz}), (\ref{Hcond}).

 Specifically, there
is always some unbroken $N=1$ supersymmetry for $PQ \neq  0$ extreme
black holes (one quarter of $N=4$ supersymmetry).
For example, for $P>0, Q>0$  the unbroken $N=1$ supersymmetry
for the solution (\ref{anz}), (\ref{Hcond}) is one combination
of third and fourth supersymmetry, $ \epsilon^{34}_{+},
\epsilon_{34}^{+}$, in the notation of ref.\cite{US}, the first
and the second being broken.
The space-time dependence of the Killing spinor in the canonical
geometry (\ref{anz}) is given by
$\epsilon = e^{\frac{1}{2} U} \epsilon_0 \,$
where $\epsilon_0$ is a constant spinor.

Consider  the extreme $PQ \not= 0$ dilatonic black holes near the
horizon,
 i.e. in the limit $\rho \rightarrow 0$.  The metric  in (\ref{anz})
becomes
\begin{equation} ds^2 =
\frac{{\rho}^2}{(M^2 - \Sigma^2)} \,  dt^2 -\frac{(M^2 -
\Sigma^2)}{\rho^2}
 \, d \rho^2 - (M^2 - \Sigma^2) \, d\Omega^2  \ . \label{RBtype}
\end{equation}
This metric is precisely the Robinson-Bertotti metric (\ref{BR})
familiar from $N=2$ supergravity  \cite{Gibbons}.
The mass parameter of the Robinson-Bertotti metric is in this case
\begin{equation}
M_{RB} = \sqrt{M^2 - \Sigma^2} = \sqrt{2|PQ|} \, \, .
\end{equation}
 The dilaton for these solutions behaves as
\begin{equation} \label{dilrbl}
{\mbox{e}}^{2 \phi} = {\mbox{e}}^{2 \phi_0} \,  \left|  \frac{P}{Q}
\right| \,
 \biggl( 1 -   {\frac{{\sqrt{2}}\, \, \Sigma}{M_{RB}^2}} \,  \rho
+ O(\rho^2) \biggr) \, ,
\end{equation}
so we see that the term linear in $\rho$ is proportional to the
dilaton charge
$\Sigma$.  The electric and magnetic fields are given by
\begin{equation}\label{FGrbl} F = {\mbox{e}}^{\phi_0} \, \frac{1}{2
Q}  \,
d\rho
\wedge dt \ ,  \qquad \tilde{G} = {\mbox{e}}^{-\phi_0} \, \frac{1}{2
P} \,
d\rho \wedge dt
\,  . \end{equation}

 Since the dilaton field has a
term linear in $\rho$, we will call this solution a
``Robinson-Bertotti type"
geometry.

Let us now see if, near the horizon, the extreme dilaton black holes
with $PQ
\not =0$ possess any additional supersymmetry and, if so, how the
charges
control which ones are unbroken.

First consider the dilatino transformations rules in the notation of
\cite{US},
\begin{equation} \frac{1}{2}\delta \Lambda_I = - \gamma^\mu
\epsilon_I \partial_\mu\phi + \frac{1}{\sqrt 2} \sigma^{\mu\nu}\left(
e^{-\phi}
F_{\mu\nu} \alpha_{IJ} -
e^\phi \tilde G_{\mu\nu}\beta_{IJ}\right)^-\epsilon^J = 0\ ,
\label{dilatino}\end{equation}
The first term in (\ref{dilatino}), ($-\gamma^a e_a^\mu \partial_\mu
\phi$),
involves the {\it flat space} derivative of the dilaton, which
vanishes near
the
horizon $\rho=0$
\begin{equation}\label{flat0}
e_a^\mu \partial_\mu \phi \sim \rho \rightarrow 0 \, ,
\end{equation}
where we used eqs. (\ref{dilrbl}), (\ref{RBtype}). The second term in
eq.
(\ref{dilatino}) is proportional to
\begin{equation}\label{controller}
\biggl(\alpha_{IJ} -  \mbox{sgn}(PQ) \beta_{IJ}\biggr)
\epsilon^J\ ,
\end{equation}
so we see that this time it is the sign of $PQ$ that controls which
combinations of supersymmetries are broken (or unbroken).

For positive $PQ$, the term in brackets in front of $\epsilon^J$ in
(\ref{controller})  vanishes under the condition that
\begin{equation}\label{constr}
\epsilon^1 = \epsilon^2 = \epsilon_1= \epsilon_2= 0\ ,
\end{equation}
i.e. the first and second supersymmetries are broken, and there are
no
constraints on the third and fourth supersymmetries.
For negative  $PQ$, the third and
fourth supersymmetries are broken, and the first and second
supersymmetries
are not,
\begin{equation}
\epsilon^3 = \epsilon^4 = \epsilon_3= \epsilon_4= 0\ .
\end{equation}

The next step is to investigate whether there are additional
constraints
coming from the supersymmetry transformation of the gravitino field
strength.
Consider first positive $PQ$. The variation of the first and the
second
gravitino vanishes because of eqs. (\ref{constr}). The transformation
rules
for the third and fourth gravitino field strength are similar to
those of
$N=2$ supergravity as given in eq. (\ref{gravitino}).  Therefore
there are indeed no additional constraints coming from
$\delta \Psi_{I\mu}$. There are several
remarkable properties of the dilatonic black hole near the horizon
making this
$N=2$ supersymmetry possible.  The first is that the scalar curvature
vanishes
near the horizon,
\begin{equation} R = 2 g^{\mu\nu} \partial_\mu \phi \cdot
\partial_\nu \phi = 2 g^{rr} (\partial_r \phi)^2 \sim \rho^2
\rightarrow 0\ ,
\end{equation}
so that the Weyl spinor is identical to the Ricci spinor.   The
second is that the geometry is of Robinson-Bertotti type, as noted
above, so
that
the covariant derivatives of the vector fields vanish and the Weyl
spinor is
zero.
Finally, as shown in eq. (\ref{flat0}),  the flat space derivative of
the
dilaton vanishes.

Therefore, we have the following situation near the horizon:
 for positive $PQ$ the unbroken $N=2$ supersymmetry consists of the
third
and the fourth ones, and for negative values of $PQ$ it is reverse:
the first
and
the second supersymmetries are unbroken whereas the third and the
fourth are
broken. Note that despite the doubling of unbroken supersymmetries
near the
horizon the unbroken supersymmetry never becomes equal  to the
maximal
possible one: only half of the possible $N=4$ supersymmetries are
restored.
This is different
from the classical extreme Reissner-Nordstr\"om solution  which, near
the
horizon,
restores maximal supersymmetry of this theory, namely $N=2$.

The difference with the classical extreme Reissner-Nordstr\"om
solution near
the horizon is also that a nontrivial dilaton dependence on $\rho$
exists and
is
proportional to the dilaton charge $\Sigma$, and that the  parameter
$M_{RB}^2$
in the Robinson-Bertotti  type geometry is not the square of the mass
of  the
extreme black hole but the difference $M^2 - \Sigma^2$.  It is
exactly this
parameter in front of the  angular part of the RB metric in eqs.
(\ref{BR}) and
(\ref{RBtype}) which defines the area of the horizon of RN and
dilaton black
holes. It is this parameter which is never zero for the non-trivial
classical
extreme black hole, but becomes zero for dilatonic extreme black
holes when
$ M^2 = \Sigma^2$ .  This happens in the limit when
either the electric or magnetic charge vanishes.  Such extreme black
holes have
unbroken $N=2$ supersymmetry \cite{US} and their geometry near the
horizon is
very different from the one described above where the limit to the
horizon is
taken with both electric and magnetic charges present.

\vskip 0.6 cm
4.  Here we will consider the purely magnetic $U(1)$ dilatonic black
holes.
 The electric case in the
canonical metric may be obtained trivially from it by the
replacements $P
\rightarrow Q$ and $\phi \rightarrow -\phi$.  The magnetic black hole
in the
stringy metric turns out to have more interesting properties than the
purely
electric one \cite{GHS}, \cite{US}. We use again the construction
given in
equations (\ref{anz}), (\ref{Hcond}). As explained above, this ansatz
solves
both
the $N=4$ supergravity equation of motion and the supersymmetry
equations
(\ref{kil}), leaving $N=2$ supersymmetry unbroken. For purely
magnetic extreme
black holes we have in  isotropic coordinates
\begin{equation}
H_1 =
{\mbox{e}}^{-\phi_0} \ , \qquad H_2 = {\mbox{e}}^{+\phi_0}
(1+\frac{\sqrt
{2}|P|}{\rho}) \ .
\end{equation}
The mass $M$ and dilaton charge $\Sigma$ are
related to the magnetic  charge $P$ as follows:
 \begin{equation}
M = \Sigma = \frac{|P| }{\sqrt {2 }} \, \, .
\end{equation}
The metric for the extreme magnetic
dilaton
black hole already possesses $N=2$ supersymmetry when viewed as a
bosonic
solution of $N=4,~d=4$ supergravity \cite{US}.  However, as can be
checked,
going
to the horizon does {\it not} result in additional supersymmetries.
The
geometry
near the horizon just keeps all 8 unbroken supersymmetries of the
black hole.
To
see this, let us consider the limit to the horizon of the extreme
purely
magnetic
black hole. This limit has been studied before in  \cite{GS}.
However, we will
investigate this limit by using our construction (\ref{anz}),
(\ref{Hcond}),
which
automatically solves the Killing equations for the unbroken
supersymmetry.

The metric has different behavior to that of the black
holes of the previous section due to the fact that  now $PQ=0$.
The most important difference is the fact that $g_{tt}^{-1}$
=$g_{\rho\rho}$
is linear in $\rho^{-1}$ rather then quadratic, as was the case
for classical  extreme Reissner-Nordstr\"om metric near the horizon
and the
$PQ\not =0$ dilatonic extreme black holes
considered above.  In addition,  the dilaton behaves differently. The
metric,
dilaton and vector fields near the horizon are:
\begin{eqnarray}
ds^2 &=& \frac{\rho}{\sqrt{2} |P|} dt^2 -
\frac{\sqrt{2} |P|}{\rho} ( d\rho^2 + \rho^2 d\Omega^2 )  \\
&=&   \frac{\rho}{\sqrt{2} |P|} \Bigl[ dt^2 -
\frac{2 P^2}{\rho^2} (d\rho^2 + \rho^2 d\Omega^2 ) \Bigr]    \ ,
\end{eqnarray}
\begin{equation}
{\mbox{e}}^{2\phi} = {\mbox{e}}^{2\phi_0} \,
\frac{\sqrt{2} | P | }{\rho}\ ,
\end{equation}
\begin{equation}\label{sol1}
G = P\, {\mbox{e}}^{2\phi_0}\sin \theta \, d\theta \wedge d\phi\ .
\end{equation}

Now let us make a change of variables
\begin{equation}
w = -\sqrt{2}\, P \, \mbox{ln}\rho\ ,
\end{equation}
so that the metric becomes
\begin{equation}
ds^2 = {1 \over \sqrt{2} |P|} \mbox{exp} \, ({\frac{-w}{\sqrt{2} P}})
 \Bigl[ dt^2 - dw^2 - 2 P^2 d\Omega^2 \Bigr] \ ,
\end{equation}
and the dilaton becomes
\begin{equation}\label{dilaton}
\phi = \phi_0+{\textstyle\frac{1}{2}} \mbox{ln}(\sqrt{2} |P|)+
\frac{w}{2 \sqrt{2}P} \ ,
\end{equation}
or
\begin{equation}
{\mbox{e}}^{2\phi} = \sqrt{2} |P| \, {\mbox{e}}^{2\phi_0} \,
\mbox{exp}(\frac{w}{\sqrt{2} P}) \ .
\end{equation}

Going to the stringy metric via the transformation $ds^2_{str} =
{\mbox{e}}^{2\phi} ds^2$, we obtain
\begin{equation}\label{sphere}
ds^2_{str} = dt^2 - dw^2 - 2 P^2 d\Omega^2 \ ,
\end{equation}
which is the direct product of a flat 2-d Minkowskian metric in the
coordinates
 $(t,w)$ and a transverse space of constant curvature ($1/ \sqrt{2
P^2} $).  In
these coordinates we see that the dilaton is linear.
This is a well-known result in the context of 2-d ``dilaton gravity"
\cite{GS},
\cite{AlStromBanks}.

The fact that the direct product of the linear dilaton vacuum
and a sphere of a constant curvature has 8 unbroken supersymmetries
follows simply from the fact that  this is a specific example of the
 construction (\ref{anz}), (\ref{Hcond}), which has been proved
\cite{US} to have
unbroken $N=2 \, , d=4$ supersymmetry. For positive $P$ the unbroken
supersymmetries are given by the 8 specific combinations \cite{US}
$\epsilon^{34}_+ , \epsilon_{34}^+ , \epsilon^{12}_-,
\epsilon_{12}^-$ of the original 16 supersymmetries, the
other 8 are broken.  For negative $P$ those which were broken
become unbroken and vice versa.

Let us see why there is no doubling of supersymmetries near the
horizon
this time, as different from the previous cases. Consider again the
dilatino
transformation rules (\ref{dilatino}). Now that the metric behaves
differently at the horizon, there is a contribution from the first
term as
well as from the vector field. The sum of those two contributions
 vanishes under the condition that half of the supersymmetries are
broken. The unbroken supersymmetries
 are listed above: going near the horizon does not change
the number of unbroken supersymmetries of the extreme purely magnetic
(or electric) dilatonic black hole.

As argued in \cite{GS} the evaporation of or scattering by
purely magnetic 4-d black
holes near extremality is closely related to the analogous processes
for 2-d black holes. It can be shown that the relation between the
masses of
the 2-d and 4-d black holes is
\begin{equation}\label{GS}
M_{2d} \sim \frac{M_{4d} - |z|}{|z|} \ ,
\end{equation}
where the two central charges of $N=4$ supersymmetry are
defined in \cite{US} to be $ |z_1|=|z_2|=|z|=
|P|/\sqrt{2}$. The supersymmetry bound for 4-d dilaton black holes
derived in \cite{US},
\begin{equation}
M_{4d} - |z| \geq 0 \ ,
\end{equation}
thus ensures that the mass of the 2-d black holes is non-negative,
\begin{equation}\label{posit}
M_{2d} \geq 0 \ .
\end{equation}
The saturation of the bound in 4-d takes place when the mass
reaches the value of the central charge,
$M_{4d} = |z| $,  $N=2$ supersymmetry becomes restored
and the evaporation stops. In 2-d this corresponds to total
evaporation of the
black hole $M_{2d} = 0$, i.e. to the linear dilaton vacuum.

Thus we conclude that stringy 2-d black holes, which are related to
 4-d black holes as in eqs. (\ref{GS}) - (\ref{posit}), may have
better
control over quantum corrections and stability of the extreme
solution due to the existence of 8 unbroken
supersymmetries and the corresponding non-renormalization theorems
\cite{US}. However, the role of an additional 2-sphere
of a radius $2 |z|$, which is required in
the geometry (\ref{sphere}), (\ref{dilaton})
to keep supersymmetry unbroken, is still to be understood.
In particular, a supersymmetric embedding of 2-d dilaton gravity may
be
looked for, which has  the linear dilaton vacuum with flat
two-dimensional
metric
as the solution with  unbroken  supersymmetry \cite{S}. This
supersymmetry will take place in a purely two-dimensional theory
without an
additional 2-sphere. This kind of supersymmetry may provide
constraints on the quantum theory in $d=2$ and be useful in the
context of a
two-dimensional toy model of quantum gravity. However, if one
considers
investigation of 2-d black holes as a way to get
insight into a more complicated (but more realistic) theory  of 4-d
black
holes,
one should remember that  physics of the 4-dimensional black
holes near the horizon is effectively two-dimensional and described
by 2-d
effective Lagrangian only in the low energy limit. It is not clear
whether
 2-d black holes may have all 8 supersymmetries possessed by 4-d
black holes
near the horizon. Therefore the problem of stable remnants  may have
quite
different solutions, depending on whether the geometry is purely
two-dimensional or four-dimensional.

\vskip 1 cm

The authors wish to thank D. Brill, G. Horowitz, A. Linde, M. Perry,
K. Stelle,
A.
Strominger and L. Susskind for useful discussions.
One of us (R.K.) would like to thank the Aspen Center for Physics,
 and participants of the Black Hole
workshop for stimulating conversations.

The work of R.K. and A.P.  was supported by NSF grant PHY-8612280.
The work of
R.K.  was supported in part
 by the John and Claire Radway Fellowship in the School of
Humanities and Sciences at  Stanford University.

\end{document}